\documentclass[aps,twocolumn,showpacs,preprintnumbers,nofootinbib,prl,superscriptaddress,10pt]{revtex4-1}

\makeatletter
\def\l@subsubsection#1#2{}
\def\l@subsubsubsection#1#2{}
\makeatother

\usepackage{graphicx,amssymb,amsmath,amsthm,amsfonts,epsfig,epsf,fixmath}
\usepackage[usenames]{color}
\usepackage{epstopdf}

\usepackage{aas_macros}
\usepackage{bm}
\usepackage{dcolumn}
\usepackage{latexsym}
\usepackage{rotating}
\usepackage{longtable}

\usepackage{enumerate}
\usepackage{tensor,multirow}
\usepackage{url}
\usepackage[linktocpage]{hyperref}

\newcommand{\tn}{\textnormal}

\begin{document}
\title{Detecting scalar fields with Extreme Mass Ratio Inspirals}

\author{Andrea Maselli}
\affiliation{Dipartimento di Fisica, ``Sapienza'' Universit\`a di Roma, Piazzale 
Aldo Moro 5, 00185, Roma, Italy}

\author{Nicola Franchini}
\affiliation{SISSA, Via Bonomea 265, 34136 Trieste, Italy and INFN Sezione di Trieste}
 \affiliation{IFPU - Institute for Fundamental Physics of the Universe, Via Beirut 2, 34014 Trieste, Italy}

\author{Leonardo Gualtieri}
\affiliation{Dipartimento di Fisica, ``Sapienza'' Universit\`a di Roma, Piazzale 
Aldo Moro 5, 00185, Roma, Italy}

\author{Thomas P.~Sotiriou}
\affiliation{School of Mathematical Sciences \& School of Physics and Astronomy, University of Nottingham, University Park, Nottingham, NG7 2RD, UK}

\begin{abstract} 
We study Extreme Mass Ratio Inspirals (EMRIs), during which a small body spirals into a 
supermassive black hole, in gravity theories with additional scalar fields. We first argue that 
no-hair theorems and the properties of known theories that manage to circumvent them 
introduce a drastic simplification to the problem: the effects of the scalar on supermassive 
black holes, if any, are mostly negligible for EMRIs in vast classes of theories. We then exploit 
this simplification to model the inspiral perturbatively and we demonstrate that the scalar 
charge of the small body leaves a significant imprint on gravitational wave emission. 
Although much higher precision is needed for waveform modelling, our results strongly 
suggest that this imprint is observable with LISA, rendering EMRIs promising probes of 
scalar fields. 
\end{abstract}

\maketitle

\noindent{{\bf{\em Introduction.}}}
The existence of additional gravitational wave (GW) polarizations with respect to general relativity (GR) is a generic
feature of alternative theories of gravity. Direct observation of these extra polaritazions would be quite challenging,
because they are expected to couple very weakly to detectors. Nonetheless, if they exist then they do affect the
emission: any extra polarization is an additional channel for energy loss for a binary system. The latter generically
loses energy at a different rate than in GR. This modifies the orbital dynamics and the GW frequency's evolution,
leaving an imprint on standard polarizations.

For comparable mass binaries in the inspiral phase, the leading-order effect comes from dipolar
emission~\cite{1975ApJ...196L..59E,Berti:2004bd,Will:2005va}. The theory has an additional field and compact objects are
``dressed'' by it. One can think of them as carrying a ``charge'' -- we use the term colloquially as we are not
necessarily 
referring to gauge fields.  Hence, to leading order each of the members of the binary acts like a monopole in the new
field and the orbiting pair emits dipole radiation. The rate of emission, namely the energy loss, depends on how much
``charge'' the compact objects carry, and more specifically on the difference between charges (or sensitivities in post-Newtonian jargon~\cite{1975ApJ...196L..59E}).  Since the effects that are associated with the additional energy loss are
cumulative, observing a long inspiral can lead to significant constraints on dipolar emission, assuming that the new
field is massless or sufficiently light, so that the corresponding interaction has sufficiently long range.\footnote{A
  large mass would make the interaction short range, quenching the emission at large
  separations~\cite{Alsing:2011er,Berti:2012bp,Ramazanoglu:2016kul}.}

Dipole emission, or more precisely absence thereof, has yielded strong constraints on massless scalars using binary
pulsar observations~\cite{Freire:2012mg}.
GW observations of binary neutrons star inspirals can significantly improve these
  constraints~\cite{Barausse:2016eii}. Moreover, they can
probe the same effect at smaller separations and in principle detect or constrain 
more massive scalars. This is a major goal for ground-based detectors
\cite{Barack:2018yly}.

Can extreme mass ratio inspirals (EMRIs) onto supermassive black holes (BHs), which will be prime targets from LISA
\cite{Audley:2017drz,Barausse:2020rsu}, yield comparable constraints?  No-hair theorems dictate that stationary BHs in
most scalar-tensor theories will just be described by the Kerr metric~\cite{chase1970event,Bekenstein:1995un,
  Hawking:1972qk, Sotiriou:2011dz,Hui:2012qt}.  Evading these theorems requires coupling the scalar to higher-order
curvature invariants~\cite{Sotiriou:2013qea,Silva:2017uqg,Doneva:2017bvd,Antoniou:2017acq}.  Indeed, the known BH
solutions with scalar hair, {\em
  e.g.}~\cite{Campbell:1991kz,Mignemi:1992nt,Kanti:1995vq,Yunes:2011we,Kleihaus:2011tg,Sotiriou:2013qea,%
  Sotiriou:2014pfa,Herdeiro:2014goa,Antoniou:2017acq,Doneva:2017bvd,Silva:2017uqg,Antoniou:2017hxj}, tend to have scalar
``charges'' that are not independent and are controlled by the mass of the BH. The more massive a BH is, the more weakly
charged it is. This is because the ``charge'' is controlled by curvature and the curvature near the horizon tends to grow
as the mass decreases.  It is then tempting to conclude that systems that involve supermassive BHs will exhibit much
smaller deviation from GR and hence will be less suitable for inspiral tests.

While this is true for comparable mass BH binaries (see {\em e.g.}~Ref.~\cite{Witek:2018dmd}), it is incorrect in
general. Consider an EMRI. So long as the companion carries a significant ``charge'', there should be emission in the
corresponding polarisation. As we will see in more detail shortly, the fact that the supermassive BH in an EMRI carries
no or very little charge is in fact a blessing in disguise from a technical perspective. Intuitively, the setup is not
much different from an accelerated electric charge. One can think of the companion as a scalar monopole that is
accelerated gravitationally by the supermassive BH and thus emits scalar (mostly dipolar) radiation. The main goal of
this paper is to demonstrate that this emission has a very significant, cumulative effect during the long inspiral of an
EMRI, which appears to be detectable by space interferometers as LISA (for similar computations for a specific class of
scalar-tensor theories see~\cite{Yunes:2011aa,Pani:2011xj}). Moreover, we shall show that the additional energy loss in
an EMRI -- and thus the dephasing of the gravitational waveform -- mainly depends on the scalar charge of the object
orbiting around the supermassive BH and has negligible dependence on other features of the underlying gravity
theory. This makes EMRIs powerful tools for tests of gravity.

\noindent{{\bf{\em General setup.}}}
To demonstrate this concretely, we start from the following action 
\begin{equation}
  S[\mathbf{g},\varphi,\Psi]=S_0[\mathbf{g},\varphi]+
 \alpha S_c[\mathbf{g},\varphi]+S_{\rm m}[\mathbf{g},\varphi,\Psi]\,,\label{action}
\end{equation}
where
\begin{equation}
  \label{S_0}
S_0= \int d^4x \frac{\sqrt{-g}}{16\pi}\left(R-\frac{1}{2}\partial_\mu\varphi\partial^\mu\varphi\right)\,,
\end{equation}
$R$ is the Ricci scalar, $\varphi$ is a scalar field, and we use units in which $G=c=1$; Greek indices
run from $0$ to $3$, while Latin indices run from $1$ to $3$.  $\alpha S_c$ describes nonminimal couplings between the
metric tensor $\mathbf{g}$ and $\varphi$, and $\alpha$ is a coupling constant with dimensions
$[\alpha]=(\rm{mass})^n$. $S_{\rm m}$ is the action of the matter fields $\Psi$.

We consider the inspiral of a body (the ``particle'') with mass $m_{\rm p}$ onto a BH of mass $M$. Since the inspiral is
an EMRI, we assume that $m_{\rm p}\ll M$. We use the so-called ``skeletonized
approach''~\cite{1975ApJ...196L..59E,Damour:1992we,Julie:2017ucp,Julie:2017rpw}, in which an extended body is treated as
a point particle, replacing the matter action $S_{\rm m}$ with
\begin{equation}
  \label{skeleton_action}
  S_{\rm p}=-\int m(\varphi)ds=-\int m(\varphi)
  \sqrt{g_{\mu\nu}\frac{dy_{\rm p}^\mu}{d\lambda}\frac{dy_{\rm p}^\nu}{d\lambda}}d\lambda\,.
\end{equation}
Here $y_{\rm p}^\mu(\lambda)$ is the worldline of (the center of mass of) the particle in a given coordinate frame, and
$m(\varphi)$ is a scalar function that depends on the value of the scalar field at the location of the particle.

In this approach it is assumed that the ``skeletonized'' body has a characteristic length-scale, $l$, which is much
smaller that the length-scale $L$ of the exterior spacetime, i.e. of the spacetime solution of the field equations in
the absence of that body. The region of spacetime in which the gravitational field of the body is large is a world-tube
with size $\sim l$, and can be treated as a worldline $y_{\rm p}^\mu(\lambda)$ in the exterior spacetime. The action
$S_{\rm p}$ is obtained by integrating the matter action $S_{\rm m}$ over this world-tube. In the case of an EMRI, the
skeletonized body and the exterior spacetime coincide with the ``particle'' of mass $m_{\rm p}$ and with the BH of mass
$M$, respectively. 

Let us consider the exterior spacetime. We assume that (perturbed) BHs in the theory under consideration are
continuously connected to the corresponding GR solution as $\alpha\rightarrow0$, and that $S_{\rm c}$ is analytic in
$\varphi$.  We identify two distinct cases in which one can describe an EMRI as the motion of a particle, described by
the skeletonized action $S_{\rm p}$ given in Eq.~\eqref{skeleton_action}, in the Kerr spacetime.

{\em Case 1:} The theory described by \eqref{action} satisfies a no-hair theorem~\cite{chase1970event,Bekenstein:1995un,
  Hawking:1972qk, Sotiriou:2011dz,Hui:2012qt,Sotiriou:2013qea,Silva:2017uqg} and, hence, 
  stationary BHs are described by the Kerr metric.

{\em Case 2:} The theory evades no-hair theorems but the coupling constant $\alpha$ is dimensionful, with $n\ge1$. Our
assumption that the BH spacetime is continuously connected to the Kerr spacetime as $\alpha\to 0$ and the fact that the
only dimensionful scale of the Kerr metric is its mass $M$, imply that any correction must depend on
\begin{equation}
  \zeta\equiv\frac{\alpha}{M^n}=q^n\frac{\alpha}{m_{\rm p}^n}\,,
\end{equation}
where $q=m_{\rm p}/M\ll 1$. Assuming that $\alpha/m_{\rm p}^n<1$ (otherwise the modifications to GR would be too large
to be consistent with current astrophysical observations \cite{Nair:2019iur}), it follows that $\zeta\ll1$, being
suppressed by the mass ratio, and thus the exterior spacetime can be approximated by the Kerr metric.

It should be stressed that Case 1 covers very wide classes of scalar-tensor theories. The theories that instead are
known to evade no-hair theorems tend to belong to Case 2. A notable example is scalar Gauss-Bonnet (sGB)
gravity~\cite{Campbell:1991kz,Mignemi:1992nt,Kanti:1995vq,Yunes:2011we,Kleihaus:2011tg,Sotiriou:2013qea,%
  Sotiriou:2014pfa,Antoniou:2017acq,Doneva:2017bvd,Silva:2017uqg}, for which
\begin{equation}
S=S_0+\frac{\alpha}{4}\int d^4x\frac{\sqrt{-g}}{16\pi}f(\varphi) {\cal G}+S_{\rm m}\,,\label{actionGB}
\end{equation}
$n=2$ and $f(\varphi)$ is a general function of the scalar field, specifying the coupling between the scalar field and
the Gauss-Bonnet invariant ${\cal G}=R^2-4R_{\mu\nu}R^{\mu\nu}+R_{\mu\nu\alpha\beta}R^{\mu\nu\alpha\beta}$.  For
massless scalars, which are expected to respect shift symmetry, this coupling is essential for evading no-hair
theorems~\cite{Sotiriou:2013qea,Saravani:2019xwx}.
    However, action \eqref{action} is far more generic. For instance, it includes any theory in which a (pseudo)scalar couples to curvature invariants ({e.g.}~generalized scalar-tensor theories or dynamical Chern-Simons gravity~\cite{Alexander:2009tp}). The analysis above can
straightforwadly be extended to multiple coupling constants with different dimensions, and to theories with more than
one scalar field~\cite{Damour:1992we,Horbatsch:2015bua,Yazadjiev:2019oul}.  The only crucial assumption is that of
continuous connection to GR BHs as the new couplings tend to zero.  

Hence, for theories falling under Cases 1 and 2, one can describe an EMRI as the motion of a particle, described by the
skeletonized action $S_{\rm p}$ given in Eq.~\eqref{skeleton_action}, in the Kerr spacetime.  This motion can be studied
using spacetime perturbation theory, i.e. expanding the field equations in the mass ratio $q\ll1$. Remarkably, the GR
modifications affect the motion of the particle, but they {\it do not} affect the background spacetime. This results in
a great simplification of the EMRI modelling, and, as we show below, it allows to make rather generic predictions of the
corresponding phenomenology.

For the rest of this Letter we shall assume, for simplicity, that the BH with mass $M$ is non-rotating, and thus
described by the Schwarzschild metric. The case of a rotating BH will be studied in a forthcoming publication.

\noindent{{\bf{\em Field Equations.}}}
Varying the action with respect to the metric tensor we obtain the following modified Einstein equations:
\begin{equation}
  G_{\mu\nu}=R_{\mu\nu}-\frac{1}{2}g_{\mu\nu}R= T^{{\rm scal}}_{\mu\nu} +
  \alpha T^{c}_{\mu\nu}+ T^{\rm p}_{\mu\nu}\,,\label{eqmetric}
\end{equation}
where $T^{{\rm scal}}_{\mu\nu}=\frac{1}{2}\partial_\mu\varphi\partial_\nu\varphi
-\frac{1}{4}g_{\mu\nu}(\partial \varphi)^2$ is the stress-energy tensor of the 
scalar field, and
\begin{equation}
T^{c}_{\mu\nu}=-\frac{16\pi}{\sqrt{-g}}\frac{\delta S_c}{\delta g^{\mu\nu}}
\end{equation}
is the stress-energy tensor associated to the coupling between gravity and 
the scalar field.
Finally,
\begin{align}
  T^{{\rm p}\,\alpha\beta}
  &=8\pi
  \int  m(\varphi)\frac{\delta^{(4)}(x-y_{p}(\lambda))}{\sqrt{-g}}
  \frac{dy_p^\alpha}{d\lambda}\frac{dy_p^\beta}{d\lambda} d\lambda\,\label{def:sourceT}
\end{align}
is the particle's stress-energy tensor. 

Variation with respect to the scalar field yields:
\begin{equation}
  \square\varphi+\frac{8\pi\alpha}{\sqrt{-g}}\frac{\delta S_c}{\delta \varphi}
  =16\pi \int m'(\varphi)
        \frac{\delta^{(4)}(x-y_{p}(\lambda))}{\sqrt{-g}}d\lambda
\label{eqscalr}
\end{equation}
and $m'(\varphi)=dm(\varphi)/d\varphi$.

In our units $[S_0]=($mass$)^2$, $[S_c]=({\rm mass})^{2-n}$.
In an EMRI, $S_c$ is evaluated on the background of the large, stationary BH, and since
the only dimensionful scale in this background is  the BH mass $M$, we expect that $S_c\sim M^{-n}S_0$.
Therefore, $\alpha T^{c}_{\mu\nu}\sim \zeta G_{\mu\nu}\ll G_{\mu\nu}$ and $\alpha\frac{\delta S_c}{\delta
  \varphi}\sim\zeta \Box\varphi\ll\Box\varphi$. 
  For an EMRI around a GR BH, the external scalar field has to be a constant, $\varphi_0$. Indeed, under our
assumptions and without the contribution of the particle the field equations \eqref{eqmetric}, \eqref{eqscalr} coincide
to those of GR with a free scalar field, for which the no-hair theorem applies.  $T^{\rm scal}_{\mu\nu}$ is quadratic in
perturbations around $\varphi=\varphi_0$ and can also be neglected.   Since
$S_c$ is analytical in $\varphi$, $\alpha T^{c}_{\mu\nu}\sim \zeta^2 G_{\mu\nu}$ and the
corrections to the background metric due to the scalar field are of order $\sim
\zeta^2$, i.e. they are suppressed at least by a factor $q^{2n}$ with respect to the leading term.
These terms can then be neglected with respect to the ``particle'' terms.

 Let us now consider Eq.~\eqref{eqscalr} in a
``buffer'' region close enough to the body to be inside the world-tube, but far-away enough to have a metric which can
be written as a perturbation of flat spacetime.  In this region, since the coupling term is negligible the scalar field
equation takes the form $\square\varphi=0$. Hence, in a reference frame $\{{\tilde x}^\mu\}$ centered on the body, its
solution has the simple form
\begin{equation}
\varphi=\varphi_0+\frac{m_{\rm p}\,d}{\tilde r}+O\left(\frac{m_{\rm p}^2}{{\tilde r}^2}\right)\label{sol:scalhom}
\end{equation}
where $d$ is the {\it dimensionless scalar charge} of the body with mass $m_{\rm p}$.  At $\tilde{r}\gg m_{\rm p}$ the
solution~\eqref{sol:scalhom} tends to the asymptotic value $\varphi_0$, which is also the value of the external scalar
field near the location of the worldtube; thus, in the particle action~\eqref{skeleton_action} (and in the source terms)
the scalar function $m(\varphi)$ and its derivative should be evaluated at $\varphi=\varphi_0$.

The value of $\varphi_0$ is determined by asymptotics, so in a realistic scenario it is fixed by cosmological
considerations and it will be theory dependent. However, it turns out to be irrelevant for our analysis. For convenience
we set $\varphi_0=0$, which amounts to the redefinition $\varphi\to \varphi -\varphi_0$ in equations \eqref{eqmetric},
\eqref{eqscalr}.

Replacing the expression in Eq.~\eqref{sol:scalhom} into Eq.~\eqref{eqscalr} after our approximations yields the
relation
$ m'(0)/m_{\rm p}=-d/4$.
Finally, in the weak-field limit the $(tt)$-component of the particle's stress-energy tensor reduces to the 
matter density of the particle $\rho=m_{\rm p}\delta^{(3)}\left(x^{i}-y^{i}_{\rm p}(\lambda)\right)$, and 
since (see Eq.~\eqref{def:sourceT})
\begin{equation}
T^{{\rm p}\,tt}=8\pi m(0)\delta^{(3)}\left(x^{i}-y^{i}_{\rm p}(\lambda)\right)+O\left(\frac{m_{\rm p}}{\tilde{r}}\right)\,,
\end{equation}
we also have $m(0)=m_{\rm p}$.  We can conclude that in the class of theories considered in this paper, 
the perturbed Einstein's equations and scalar field equations for EMRIs have the form
\begin{align}
G_{\mu\nu}&=T^{\rm p}_{\mu\nu}=8\pi m_{\rm p}\int \frac{\delta^{(4)}(x-y_{p}(\lambda))}{\sqrt{-g}}
\frac{dy_p^\alpha}{d\lambda}\frac{dy_p^\beta}{d\lambda} d\lambda\label{eq:pertG}\\
  \square\varphi&=-4\pi d\,m_{\rm p}\int \frac{\delta^{(4)}(x-y_{p}(\lambda))}{\sqrt{-g}}d\lambda\,.\label{eq:pertphi}
\end{align}
While the perturbed Einstein's equations~\eqref{eq:pertG} coincide with the corresponding 
equations in GR, the perturbed scalar field equations~\eqref{eq:pertphi} have a source term 
which is proportional to the scalar charge $d$. 
All information about the underlying gravity theory is encoded in the scalar
  charge $d$, which thus universally captures the changes in the EMRI dynamics.

\noindent{{\bf{\em Perturbations.}}}
To study EMRI's evolution in theories specified by Eq.~\eqref{action} and belonging to the Cases 1 and
2 discussed above, we compute the perturbations around a Schwarzschild BH induced by a particle with mass $m_{\rm p}$
which takes into account beyond-GR corrections in the source term. We consider linear order perturbations to the
gravitational and the scalar sector, i.e. we expand both the metric tensor and $\varphi$ around a background,
$g_{\alpha\beta}=g^0_{\alpha\beta}+h_{\alpha\beta}$ and $\varphi=\varphi_0+\varphi_1$, where $g^0_{\alpha\beta}$
describes the Schwarzschild spacetime (in Schwarzschild coordinates $(t,r,\theta,\phi)$), and -- as discussed above --
$\varphi_0=0$. We decompose $h_{\mu\nu}$ and $\varphi_1$ in tensor and scalar spherical harmonics, respectively. The
metric perturbations decouple in two classes, known as {\it polar} and {\it axial} perturbations,
$h_{\alpha\beta}=h^\tn{pol}_{\alpha\beta}+h^\tn{ax}_{\alpha\beta}$, according to their properties under parity
transformation \cite{Regge:1957td,Zerilli:1971wd,Zerilli:1970se}.  For binaries in circular orbits both sectors are
excited.  In the Schwarzschild background, the metric and the scalar field perturbations are decoupled. In this set-up,
and working within the so-called Regge-Wheeler gauge, the components of the metric perturbations $h^\tn{pol}_{\mu\nu}$ $(h^\tn{ax}_{\mu\nu})$ reduce to a single
function $Z_{\ell m}$ ($R_{\ell m}$). In the frequency domain tensor and scalar perturbations satisfy the wave equation
\begin{equation}
\frac{d^2 \psi_{\ell m}}{dr_\star^2}+\left(\omega^2-e^{-\lambda} V\right)\psi_{\ell m}=J\,,
\label{master}
\end{equation}
where $e^{-\lambda}=1-\frac{2M}{r}$, $\psi_{\ell m}=(Z_{\ell m},R_{\ell m},
\delta\varphi_{\ell m})$, $r_\star=r+2M\log(r/2M-1)$ is the tortoise coordinate, 
$V$ is a $3\times3$ diagonal matrix with
\begin{equation}
V_{11}=2\frac{9M^3+9M^2r\Lambda+3Mr^2\Lambda^2+r^3\Lambda^2(1+\Lambda)}{r^3(3M+r\Lambda)}\,,
\end{equation}
$V_{22}=\frac{\ell(\ell+1)}{r^2}-\frac{6M}{r^3}$, $V_{33}=\frac{\ell(\ell+1)}{r^2}+\frac{2M}{r^3}$ and
$\Lambda=\ell(\ell+1)/2-1$. The source's terms $J_Z,J_R$ 
are explicitly given in
\cite{Sago:2002fe}, while the scalar field component reads:
\begin{equation}
J_\varphi=-d\,m_{\rm p}\frac{4\pi P_{\ell m}(\frac{\pi}{2})}{r^{3/2}e^{\lambda}}\sqrt{r-3M} \delta(r-r_{\rm p})
\delta(\omega-m\omega_{\rm p})\,.
\end{equation}
Here $r_{\rm p}$ is the particle's coordinate radius, $\omega_{\rm p}=(M/r_p^3)^{1/2}$ and $P_{\ell
  m}(\theta)$ the Legendre polynomials.
  
We numerically integrate the wave equations~\eqref{master} by first finding the homogeneous solutions at the horizon
$\psi_{\ell m}^{(-)}$ and at infinity $\psi_{\ell m}^{(+)}$, which satisfy the boundary condition of purely ingoing and
outgoing waves, respectively, i.e. $\psi_{\ell m}^{(\pm)}\sim e^{\mp i\omega r_\star}$.  The non-homogeneous solution
$\psi_{\ell m}(r_\star)$ is obtained by integrating the homogenous part over the source terms.  Evaluating the solution
at the horizon and at infinity we get
\begin{equation}
  \psi_{\ell m}^{\pm}\equiv
  \lim_{r_\star\rightarrow\pm\infty}\psi_{\ell m}(r_*)
  =e^{\pm i\omega r_\star}\int_{2M}^\infty\frac{\psi^{(\mp)}_{\ell m} J}{W}dr_\star\,,\label{def:green}
\end{equation}
where $W=\psi^{'(+)}_{\ell m}\psi^{(-)}_{\ell m}-\psi^{'(-)}_{\ell m}\psi^{(+)}_{\ell m}$ 
is the Wronskian and the prime denotes derivative with respect to $r_\star$. 
From Eqns.~\eqref{def:green} we can compute the gravitational 
and scalar energy flux at the horizon and at infinity \cite{Martel:2003jj,Blazquez-Salcedo:2016enn}:
\begin{align}
  \dot{E}^{\pm}_{\rm grav}&=\frac{1}{64\pi}\sum_{\ell=2}^\infty\sum_{m=-\ell}^\ell
  \frac{(\ell+2)!}{(\ell-2)!}(\omega^2\vert Z^{\pm}_{\ell m}\vert^2+4\vert R^{\pm}_{\ell m}\vert^2)\,,\nonumber\\ 
  \dot{E}^{\pm}_{\rm scal}&=\frac{1}{32\pi}\sum_{\ell=1}^\infty\sum_{m=-\ell}^\ell
  \omega^2\vert \delta\varphi^{\pm}_{\ell m}\vert^2\ .\label{escal}
\end{align}

\begin{figure}[!htbp]
\centering
\includegraphics[width=0.5\textwidth]{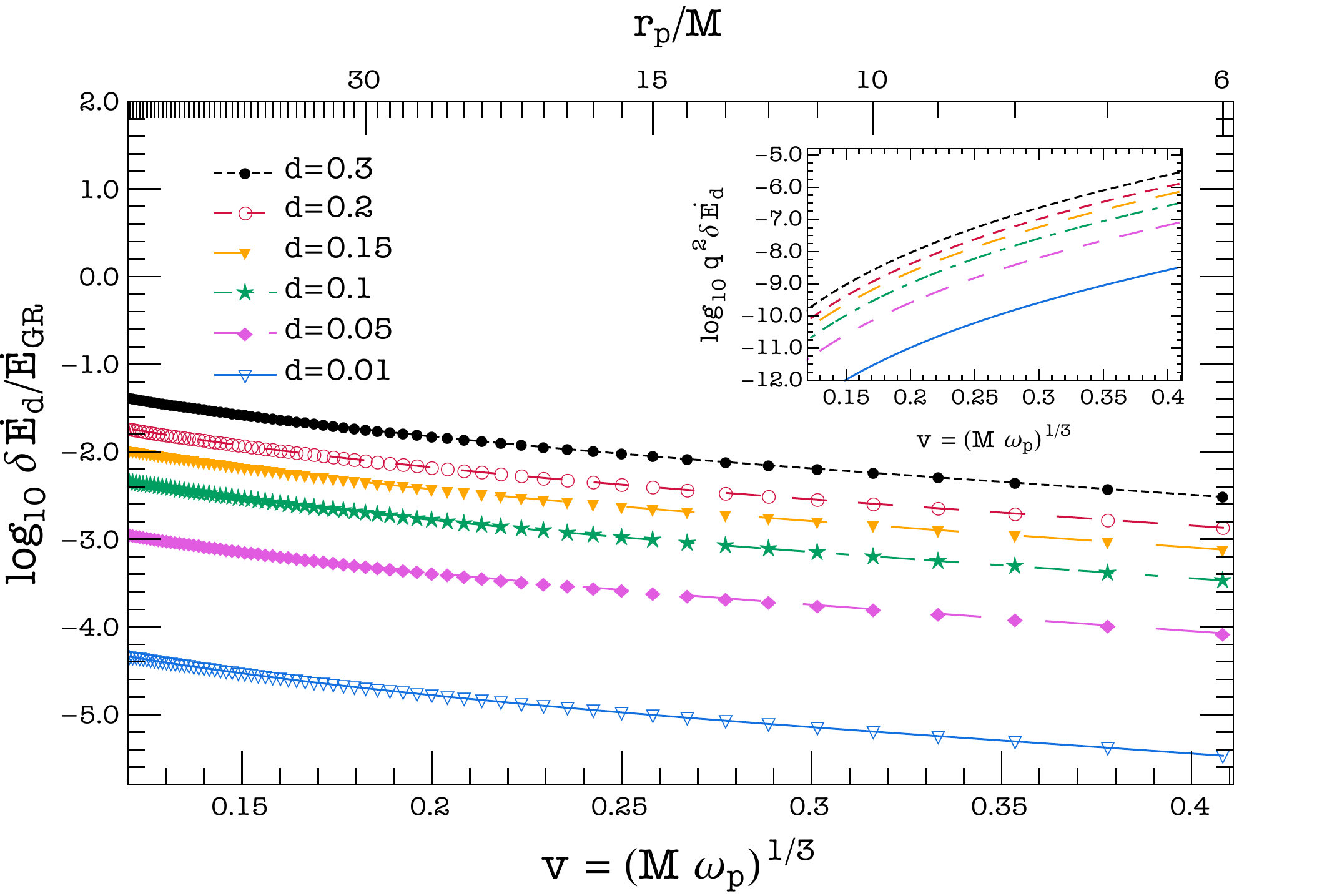}
\caption{Relative difference between the GW flux in modified gravity and in GR as a function of the orbital velocity
  $v=(M\omega_{p})^{1/3}$ (or radius $r_p/M$), and of the scalar charge $d$. The inset
  shows the values of $q^2\delta\dot{E}_d$.}\label{fig:flux_circ}
\end{figure}

\noindent{{\bf{\em Results.}}}
We compute the the total energy flux 
\begin{equation}
\dot{E}=\dot{E}^{+}_{\rm grav}+\dot{E}^{-}_{\rm grav}+\dot{E}^{+}_{\rm scal}+\dot{E}^{-}_{\rm scal}=\dot{E}_\tn{GR}+\delta\dot{E}_d
\end{equation}
summing all the multipole contributions up to $\ell=5$, where $\dot{E}_\tn{GR}$ is the energy flux emitted in GR by a
binary system with the same masses $m_{\rm p}$, $M$. Since, as discussed above, the perturbed Einstein's equations
coincide with the corresponding equations in GR, the correction to the energy flux is only due to the scalar field
emission at infinity and at the horizon, $\delta\dot{E}_d=\dot{E}^{+}_{\rm scal}+\dot{E}^{-}_{\rm scal}$.  Figure
\ref{fig:flux_circ} shows the relative correction $\delta \dot{E}_d/\dot{E}_\tn{GR}$ as a function of the orbital
velocity $v=(M\omega_\tn{p})^{1/3}$, while the inset provides the value of $\delta\dot{E}_d$.  Note that $\delta
\dot{E}_d$ formally enters at the same order in the mass ratio $q$ as the GR contribution: for a given orbital
configuration the normalized flux $q^{2}\delta\dot{E}_d$ only
depends on dimensionless scalar charge $d$. The scalar flux increases as the binary inspirals towards the ISCO at
$r=6M$, accelerating the coalescence due to the extra leakage of energy. The ratio $\delta \dot{E}_d/\dot{E}_\tn{GR}$
decreases for smaller orbital separations, since the gravitational term $\dot{E}^{\pm}_{\rm grav}$ grows faster than the
scalar field contribution at large frequencies. The relative difference between the total flux in GR and in the modified
gravity theory can be $\sim1\%$ close to the plunge.

Having computed the emitted energy flux, we can determine the EMRI's adiabatic evolution, i.e.~the GW phase $\phi$ as a
function of the frequency $f=\omega_p/\pi$:
\begin{equation}
\frac{d\phi}{df}=\frac{f}{\dot{f}}\,,\quad \dot{f}=\frac{3}{2}\frac{f}{r_p}\frac{dr}{dE_{orb}}\bigg\vert_{r_p}\dot{E}\,,
\end{equation}
with $E_{orb}$ particle's orbital energy.  The total phase can be written as $\phi(f)=\phi_\tn{GR}(f)+\delta\phi_d(f)$
where both the GR and the scalar field contribution are of the order $\mathcal{O}(1/q)$. The correction $q \delta
\phi_d$ is indeed universal, and depends only on the normalized charge $d$.

\begin{figure}[!htbp]
\centering
\includegraphics[width=0.5\textwidth]{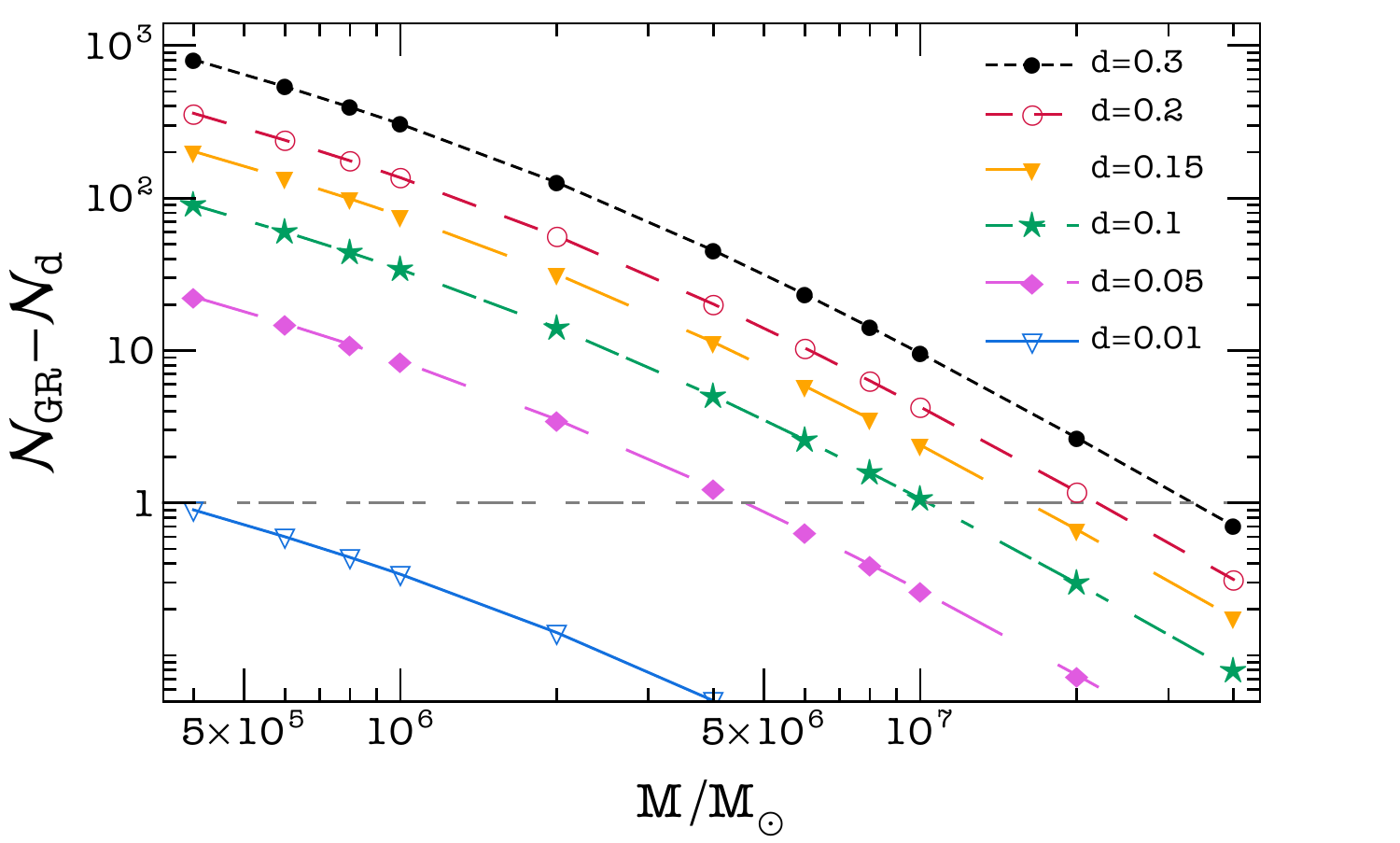}
\caption{Difference in the number of GW cycles accumulated by EMRIs on 
circular orbits with $\mu=10M_\odot$, $M\in[5\times10^5,10^8]M_\odot$, and 
different values of the dimensionless scalar charge $d$. All binaries are observed 
four years before merger.}\label{fig:dephasing_circ}
\end{figure}

To quantify the impact of GR modifications on possible GW detections by future interferometers like LISA we compute the
number of cycles accumulated before the merger \cite{Berti:2004bd}:
\begin{equation}
{\cal N}=\int_{f_{min}}^{f_{\max}}\frac{f}{\dot{f}}df\ .
\end{equation}
We choose $f_{max}=(6^{3/2}\pi M)^{-1}$ and $f_{min}=\max[f_\tn{T},10^{-4}]$, where $f_\tn{T}$ is the GW frequency $4$
years before the ISCO \cite{Pani:2011xj}, which represents the typical observing time of LISA \cite{Audley:2017drz}.
Figure \ref{fig:dephasing_circ} shows $\Delta{\cal N}={\cal N}_\tn{GR}-{\cal N}_d$ for some prototype systems with
$\mu=10M_\odot$. The difference is always positive, since the scalar field emission increases the energy loss by the
binary. $\Delta{\cal N}$ decreases monotonically as the mass of the central object grows, and it is strongly dependent
on the scalar charge. We find that for $d\gtrsim 0.01$ the dephasing can be larger than $1$ radiant (the standard conservative value for a detectable dephasing)  for $M\sim 4\cdot10^6M_\odot$.  For lighter BHs with
$M\sim 10^5M_\odot$ and large $d$, $\Delta {\cal N}$ is significantly higher and can be as large as $\sim10^3$ radians.

In principle, a potential degeneracy between the scalar charge and the black hole masses may affect a detector's ability to distinguish a non-GR signal from a GR signal with different parameters. Even if such degeneracy plagues single observables ({\em e.g.}~number of cycles), one expects it to be lifted by more detailed waveform analysis. Preliminary results in this direction confirm this expectation and a systematic analysis will be presented elsewhere \cite{degeneracies}.

\noindent{{\bf{\em A case study: sGB gravity.}}}
Figures~\ref{fig:flux_circ}, \ref{fig:dephasing_circ} show that EMRIs are a probe of the particle's scalar
charge. Typically, this quantity is related to the fundamental coupling constant $\alpha$ of the modified gravity
theory.

Let us consider, for instance, the case of sGB gravity with $f'(0)\neq0$ (i.e., excluding theories which allow for BH
scalarization~\cite{Silva:2017uqg,Doneva:2017bvd}).  If the body is a BH, its dimensionless scalar charge is
proportional to the dimensionless coupling constant of the theory $\beta\equiv q^{-2}\zeta=\alpha/m_{\rm
  p}^2$~\cite{Kanti:1995vq,Pani:2009wy,Julie:2019sab}.  The explicit form of $d(\beta)$ has been derived
in~\cite{Julie:2019sab}. Taking into account the different normalization conventions, one finds that, for instance,
$d=2\beta+\frac{73}{30}\beta^2+\frac{15577}{2520}\beta^3+O(\beta^4)$ for Einstein-dilaton Gauss-Bonnet
gravity~\cite{Kanti:1995vq,Pani:2009wy} ($f(\varphi)=e^{\varphi}$), while $d=2\beta+\frac{73}{60}\beta^3$ for
shift-symmetric sGB gravity~\cite{Sotiriou:2013qea,Sotiriou:2014pfa} ($f(\varphi)=\varphi$).

\noindent{{\bf{\em Conclusions.}}}
EMRIs are golden binaries to test fundamental physics in the strong-gravity
regime~\cite{Glampedakis:2005cf,Barack:2006pq,Barausse:2016eii,Babak:2017tow,Cardoso:2018zhm,Piovano:2020ooe}.  In
gravity theories with additional scalar fields, the enhanced energy emission 
during the inspiral leads to a cumulative dephasing of the gravitational waveform.  We 
showed that for theories satisfying no-hair theorems or having dimensionful
coupling constants, the central BH of an EMRI can be taken to be the Kerr metric, 
and the modification of the waveform only depends on the scalar charge of the 
inspiraling body. Using this significant simplification, we demonstrated that 
the corresponding dephasing should be detectable by LISA. Interestingly, the vast 
majority of known theories with additional scalar field satisfy our 
assumptions \cite{Berti:2015itd,Barack:2018yly}. Our results imply that EMRIs can 
be excellent systems for probing the existence of scalar fields and constrain 
fundamental physics. 

For any given modified gravity theory, a bound (or a measurement) of the 
scalar charge obtained from the detection of an EMRI waveform can be translated 
in a bound (or a measurement) of the fundamental coupling constant of the theory. Forecasts on scalar charge constraints for a given detector can act as a theory-independent assessment of its potential to test GR.

The approach and key simplifications we have introduced here are a critical first step towards a consistent description of EMRIs beyond GR. Our formalism can be straightforwardly extended to rotating BHs, as well
  as to generic ({\em e.g.}~eccentric) orbits, and to theories with further polarizations.  A more
  challenging problem is to study the effect of extra degrees of freedom on self-force corrections. The
  latter are essential for accurate EMRI modelling \cite{Pound:2015tma,Barack:2018yvs}, and are currently under intense
  studies aimed to provide second-order corrections to the binary dynamical evolution \cite{Pound:2019lzj}. The
  separation of scales discussed in this paper is expected to greatly reduce the complexity of self-force description
  beyond GR.

\begin{acknowledgments}

\noindent{{\bf{\em Acknowledgments.}}}
We thank Felix Juli\'e, Emanuele Berti and Susanna Barsanti 
for useful discussions. The authors would like to 
acknowledge networking support by the COST Action CA16104. A.M. acknowledge support 
from the Amaldi Research Center funded by the MIUR program "Dipartimento di Eccellenza" 
(CUP: B81I18001170001). N.F. acknowledges financial support provided under the European 
Union's H2020 ERC Consolidator Grant ``GRavity from Astrophysical to Microscopic Scales'' 
grant agreement no. GRAMS-815673. T. P. S. acknowledges partial support from the STFC 
Consolidated Grant No. ST/P000703/1.
\end{acknowledgments}

\bibliographystyle{utphys}
\bibliography{Ref}

\providecommand{\href}[2]{#2}\begingroup\raggedright\begin{thebibliography}{10}

\bibitem{1975ApJ...196L..59E}
D.~M. {Eardley}, ``{Observable effects of a scalar gravitational field in a
  binary pulsar.},'' \href{http://dx.doi.org/10.1086/181744}{{\em The
  Astrophysical Journal} {\bfseries 196} (Mar, 1975) L59--L62}.

\bibitem{Berti:2004bd}
E.~Berti, A.~Buonanno, and C.~M. Will, ``{Estimating spinning binary parameters
  and testing alternative theories of gravity with LISA},''
  \href{http://dx.doi.org/10.1103/PhysRevD.71.084025}{{\em Phys. Rev.}
  {\bfseries D71} (2005) 084025},
\href{http://arxiv.org/abs/gr-qc/0411129}{{\ttfamily arXiv:gr-qc/0411129
  [gr-qc]}}.

\bibitem{Will:2005va}
C.~M. Will, ``{The Confrontation between general relativity and experiment},''
  \href{http://dx.doi.org/10.12942/lrr-2006-3}{{\em Living Rev. Rel.}
  {\bfseries 9} (2006) 3}, \href{http://arxiv.org/abs/gr-qc/0510072}{{\ttfamily
  arXiv:gr-qc/0510072}}.

\bibitem{Alsing:2011er}
J.~Alsing, E.~Berti, C.~M. Will, and H.~Zaglauer, ``{Gravitational radiation
  from compact binary systems in the massive Brans-Dicke theory of gravity},''
  \href{http://dx.doi.org/10.1103/PhysRevD.85.064041}{{\em Phys. Rev. D}
  {\bfseries 85} (2012) 064041},
  \href{http://arxiv.org/abs/1112.4903}{{\ttfamily arXiv:1112.4903 [gr-qc]}}.

\bibitem{Berti:2012bp}
E.~Berti, L.~Gualtieri, M.~Horbatsch, and J.~Alsing, ``{Light scalar field
  constraints from gravitational-wave observations of compact binaries},''
  \href{http://dx.doi.org/10.1103/PhysRevD.85.122005}{{\em Phys. Rev. D}
  {\bfseries 85} (2012) 122005},
  \href{http://arxiv.org/abs/1204.4340}{{\ttfamily arXiv:1204.4340 [gr-qc]}}.

\bibitem{Ramazanoglu:2016kul}
F.~M. Ramazano\u~glu and F.~Pretorius, ``{Spontaneous Scalarization with
  Massive Fields},'' \href{http://dx.doi.org/10.1103/PhysRevD.93.064005}{{\em
  Phys. Rev. D} {\bfseries 93} no.~6, (2016) 064005},
  \href{http://arxiv.org/abs/1601.07475}{{\ttfamily arXiv:1601.07475 [gr-qc]}}.

\bibitem{Freire:2012mg}
P.~C. Freire, N.~Wex, G.~Esposito-Farese, J.~P. Verbiest, M.~Bailes, B.~A.
  Jacoby, M.~Kramer, I.~H. Stairs, J.~Antoniadis, and G.~H. Janssen, ``{The
  relativistic pulsar-white dwarf binary PSR J1738+0333 II. The most stringent
  test of scalar-tensor gravity},''
  \href{http://dx.doi.org/10.1111/j.1365-2966.2012.21253.x}{{\em Mon. Not. Roy.
  Astron. Soc.} {\bfseries 423} (2012) 3328},
  \href{http://arxiv.org/abs/1205.1450}{{\ttfamily arXiv:1205.1450
  [astro-ph.GA]}}.

\bibitem{Barausse:2016eii}
E.~Barausse, N.~Yunes, and K.~Chamberlain, ``{Theory-Agnostic Constraints on
  Black-Hole Dipole Radiation with Multiband Gravitational-Wave
  Astrophysics},'' \href{http://dx.doi.org/10.1103/PhysRevLett.116.241104}{{\em
  Phys. Rev. Lett.} {\bfseries 116} no.~24, (2016) 241104},
  \href{http://arxiv.org/abs/1603.04075}{{\ttfamily arXiv:1603.04075 [gr-qc]}}.

\bibitem{Barack:2018yly}
L.~Barack {\em et~al.}, ``{Black holes, gravitational waves and fundamental
  physics: a roadmap},'' \href{http://dx.doi.org/10.1088/1361-6382/ab0587}{{\em
  Class. Quant. Grav.} {\bfseries 36} no.~14, (2019) 143001},
  \href{http://arxiv.org/abs/1806.05195}{{\ttfamily arXiv:1806.05195 [gr-qc]}}.

\bibitem{Audley:2017drz}
H.~{Audley}, S.~{Babak}, J.~{Baker}, E.~{Barausse}, P.~{Bender}, E.~{Berti},
  P.~{Binetruy}, M.~{Born}, D.~{Bortoluzzi}, J.~{Camp}, C.~{Caprini},
  V.~{Cardoso}, M.~{Colpi}, J.~{Conklin}, N.~{Cornish}, C.~{Cutler}, {\em
  et~al.}, ``{Laser Interferometer Space Antenna},'' {\em ArXiv e-prints}
  (Feb., 2017) , \href{http://arxiv.org/abs/1702.00786}{{\ttfamily
  arXiv:1702.00786 [astro-ph.IM]}}.

\bibitem{Barausse:2020rsu}
E.~Barausse {\em et~al.}, ``{Prospects for Fundamental Physics with LISA},''
  \href{http://arxiv.org/abs/2001.09793}{{\ttfamily arXiv:2001.09793 [gr-qc]}}.

\bibitem{chase1970event}
J.~Chase, ``Event horizons in static scalar-vacuum space-times,'' {\em
  Communications in Mathematical Physics} {\bfseries 19} no.~4, (1970)
  276--288.

\bibitem{Bekenstein:1995un}
J.~Bekenstein, ``{Novel ``no-scalar-hair'' theorem for black holes},''
  \href{http://dx.doi.org/10.1103/PhysRevD.51.R6608}{{\em Phys. Rev. D}
  {\bfseries 51} no.~12, (1995) 6608}.

\bibitem{Hawking:1972qk}
S.~Hawking, ``{Black holes in the Brans-Dicke theory of gravitation},''
  \href{http://dx.doi.org/10.1007/BF01877518}{{\em Commun. Math. Phys.}
  {\bfseries 25} (1972) 167--171}.

\bibitem{Sotiriou:2011dz}
T.~P. Sotiriou and V.~Faraoni, ``{Black holes in scalar-tensor gravity},''
  \href{http://dx.doi.org/10.1103/PhysRevLett.108.081103}{{\em Phys.\ Rev.\
  Lett.} {\bfseries 108} (2012) 081103},
  \href{http://arxiv.org/abs/1109.6324}{{\ttfamily arXiv:1109.6324 [gr-qc]}}.

\bibitem{Hui:2012qt}
L.~Hui and A.~Nicolis, ``{No-Hair Theorem for the Galileon},''
  \href{http://dx.doi.org/10.1103/PhysRevLett.110.241104}{{\em Phys. Rev.
  Lett.} {\bfseries 110} (2013) 241104},
  \href{http://arxiv.org/abs/1202.1296}{{\ttfamily arXiv:1202.1296 [hep-th]}}.

\bibitem{Sotiriou:2013qea}
T.~P. Sotiriou and S.-Y. Zhou, ``{Black hole hair in generalized scalar-tensor
  gravity},'' \href{http://dx.doi.org/10.1103/PhysRevLett.112.251102}{{\em
  Phys.\ Rev.\ Lett.} {\bfseries 112} (2014) 251102},
  \href{http://arxiv.org/abs/1312.3622}{{\ttfamily arXiv:1312.3622 [gr-qc]}}.

\bibitem{Silva:2017uqg}
H.~O. Silva, J.~Sakstein, L.~Gualtieri, T.~P. Sotiriou, and E.~Berti,
  ``{Spontaneous scalarization of black holes and compact stars from a
  Gauss-Bonnet coupling},''
  \href{http://dx.doi.org/10.1103/PhysRevLett.120.131104}{{\em Phys. Rev.
  Lett.} {\bfseries 120} no.~13, (2018) 131104},
\href{http://arxiv.org/abs/1711.02080}{{\ttfamily arXiv:1711.02080 [gr-qc]}}.

\bibitem{Doneva:2017bvd}
D.~D. Doneva and S.~S. Yazadjiev, ``{New Gauss-Bonnet Black Holes with
  Curvature-Induced Scalarization in Extended Scalar-Tensor Theories},''
  \href{http://dx.doi.org/10.1103/PhysRevLett.120.131103}{{\em Phys. Rev.
  Lett.} {\bfseries 120} no.~13, (2018) 131103},
\href{http://arxiv.org/abs/1711.01187}{{\ttfamily arXiv:1711.01187 [gr-qc]}}.

\bibitem{Antoniou:2017acq}
G.~Antoniou, A.~Bakopoulos, and P.~Kanti, ``{Evasion of No-Hair Theorems and
  Novel Black-Hole Solutions in Gauss-Bonnet Theories},''
  \href{http://dx.doi.org/10.1103/PhysRevLett.120.131102}{{\em Phys. Rev.
  Lett.} {\bfseries 120} no.~13, (2018) 131102},
  \href{http://arxiv.org/abs/1711.03390}{{\ttfamily arXiv:1711.03390
  [hep-th]}}.

\bibitem{Campbell:1991kz}
B.~A. Campbell, N.~Kaloper, and K.~A. Olive, ``{Classical hair for Kerr-Newman
  black holes in string gravity},''
  \href{http://dx.doi.org/10.1016/0370-2693(92)91452-F}{{\em Phys. Lett. B}
  {\bfseries 285} (1992) 199--205}.

\bibitem{Mignemi:1992nt}
S.~Mignemi and N.~Stewart, ``{Charged black holes in effective string
  theory},'' \href{http://dx.doi.org/10.1103/PhysRevD.47.5259}{{\em Phys. Rev.
  D} {\bfseries 47} (1993) 5259--5269},
  \href{http://arxiv.org/abs/hep-th/9212146}{{\ttfamily arXiv:hep-th/9212146}}.

\bibitem{Kanti:1995vq}
P.~Kanti, N.~E. Mavromatos, J.~Rizos, K.~Tamvakis, and E.~Winstanley,
  ``{Dilatonic black holes in higher curvature string gravity},''
  \href{http://dx.doi.org/10.1103/PhysRevD.54.5049}{{\em Phys. Rev.} {\bfseries
  D54} (1996) 5049--5058},
\href{http://arxiv.org/abs/hep-th/9511071}{{\ttfamily arXiv:hep-th/9511071
  [hep-th]}}.

\bibitem{Yunes:2011we}
N.~Yunes and L.~C. Stein, ``{Non-Spinning Black Holes in Alternative Theories
  of Gravity},'' \href{http://dx.doi.org/10.1103/PhysRevD.83.104002}{{\em Phys.
  Rev. D} {\bfseries 83} (2011) 104002},
  \href{http://arxiv.org/abs/1101.2921}{{\ttfamily arXiv:1101.2921 [gr-qc]}}.

\bibitem{Kleihaus:2011tg}
B.~Kleihaus, J.~Kunz, and E.~Radu, ``{Rotating Black Holes in Dilatonic
  Einstein-Gauss-Bonnet Theory},''
  \href{http://dx.doi.org/10.1103/PhysRevLett.106.151104}{{\em Phys. Rev.
  Lett.} {\bfseries 106} (2011) 151104},
  \href{http://arxiv.org/abs/1101.2868}{{\ttfamily arXiv:1101.2868 [gr-qc]}}.

\bibitem{Sotiriou:2014pfa}
T.~P. Sotiriou and S.-Y. Zhou, ``{Black hole hair in generalized scalar-tensor
  gravity: An explicit example},''
  \href{http://dx.doi.org/10.1103/PhysRevD.90.124063}{{\em Phys. Rev. D}
  {\bfseries 90} (2014) 124063},
  \href{http://arxiv.org/abs/1408.1698}{{\ttfamily arXiv:1408.1698 [gr-qc]}}.

\bibitem{Herdeiro:2014goa}
C.~A.~R. Herdeiro and E.~Radu, ``{Kerr black holes with scalar hair},''
  \href{http://dx.doi.org/10.1103/PhysRevLett.112.221101}{{\em Phys. Rev.
  Lett.} {\bfseries 112} (2014) 221101},
  \href{http://arxiv.org/abs/1403.2757}{{\ttfamily arXiv:1403.2757 [gr-qc]}}.

\bibitem{Antoniou:2017hxj}
G.~Antoniou, A.~Bakopoulos, and P.~Kanti, ``{Black-Hole Solutions with Scalar
  Hair in Einstein-Scalar-Gauss-Bonnet Theories},''
  \href{http://dx.doi.org/10.1103/PhysRevD.97.084037}{{\em Phys. Rev. D}
  {\bfseries 97} no.~8, (2018) 084037},
  \href{http://arxiv.org/abs/1711.07431}{{\ttfamily arXiv:1711.07431
  [hep-th]}}.

\bibitem{Witek:2018dmd}
H.~Witek, L.~Gualtieri, P.~Pani, and T.~P. Sotiriou, ``{Black holes and binary
  mergers in scalar Gauss-Bonnet gravity: scalar field dynamics},''
  \href{http://dx.doi.org/10.1103/PhysRevD.99.064035}{{\em Phys. Rev. D}
  {\bfseries 99} no.~6, (2019) 064035},
  \href{http://arxiv.org/abs/1810.05177}{{\ttfamily arXiv:1810.05177 [gr-qc]}}.

\bibitem{Yunes:2011aa}
N.~Yunes, P.~Pani, and V.~Cardoso, ``{Gravitational Waves from Quasicircular
  Extreme Mass-Ratio Inspirals as Probes of Scalar-Tensor Theories},''
  \href{http://dx.doi.org/10.1103/PhysRevD.85.102003}{{\em Phys.\ Rev.\ D}
  {\bfseries 85} (2012) 102003},
  \href{http://arxiv.org/abs/1112.3351}{{\ttfamily arXiv:1112.3351 [gr-qc]}}.

\bibitem{Pani:2011xj}
P.~Pani, V.~Cardoso, and L.~Gualtieri, ``{Gravitational waves from extreme
  mass-ratio inspirals in Dynamical Chern-Simons gravity},''
  \href{http://dx.doi.org/10.1103/PhysRevD.83.104048}{{\em Phys. Rev.}
  {\bfseries D83} (2011) 104048},
\href{http://arxiv.org/abs/1104.1183}{{\ttfamily arXiv:1104.1183 [gr-qc]}}.

\bibitem{Damour:1992we}
T.~Damour and G.~Esposito-Farese, ``{Tensor multiscalar theories of
  gravitation},'' \href{http://dx.doi.org/10.1088/0264-9381/9/9/015}{{\em
  Class.\ Quant.\ Grav.} {\bfseries 9} (1992) 2093--2176}.

\bibitem{Julie:2017ucp}
F.-L. Julié, ``{Reducing the two-body problem in scalar-tensor theories to the
  motion of a test particle : a scalar-tensor effective-one-body approach},''
  \href{http://dx.doi.org/10.1103/PhysRevD.97.024047}{{\em Phys.\ Rev.\ D}
  {\bfseries 97} no.~2, (2018) 024047},
  \href{http://arxiv.org/abs/1709.09742}{{\ttfamily arXiv:1709.09742 [gr-qc]}}.

\bibitem{Julie:2017rpw}
F.-L. Julié, ``{On the motion of hairy black holes in Einstein-Maxwell-dilaton
  theories},'' \href{http://dx.doi.org/10.1088/1475-7516/2018/01/026}{{\em
  JCAP} {\bfseries 01} (2018) 026},
  \href{http://arxiv.org/abs/1711.10769}{{\ttfamily arXiv:1711.10769 [gr-qc]}}.

\bibitem{Nair:2019iur}
R.~Nair, S.~Perkins, H.~O. Silva, and N.~Yunes, ``{Fundamental Physics
  Implications for Higher-Curvature Theories from Binary Black Hole Signals in
  the LIGO-Virgo Catalog GWTC-1},''
  \href{http://dx.doi.org/10.1103/PhysRevLett.123.191101}{{\em Phys. Rev.
  Lett.} {\bfseries 123} no.~19, (2019) 191101},
  \href{http://arxiv.org/abs/1905.00870}{{\ttfamily arXiv:1905.00870 [gr-qc]}}.

\bibitem{Saravani:2019xwx}
M.~Saravani and T.~P. Sotiriou, ``{Classification of shift-symmetric Horndeski
  theories and hairy black holes},''
  \href{http://dx.doi.org/10.1103/PhysRevD.99.124004}{{\em Phys. Rev. D}
  {\bfseries 99} no.~12, (2019) 124004},
  \href{http://arxiv.org/abs/1903.02055}{{\ttfamily arXiv:1903.02055 [gr-qc]}}.

\bibitem{Alexander:2009tp}
S.~Alexander and N.~Yunes, ``{Chern-Simons Modified General Relativity},''
  \href{http://dx.doi.org/10.1016/j.physrep.2009.07.002}{{\em Phys. Rept.}
  {\bfseries 480} (2009) 1--55},
  \href{http://arxiv.org/abs/0907.2562}{{\ttfamily arXiv:0907.2562 [hep-th]}}.

\bibitem{Horbatsch:2015bua}
M.~Horbatsch, H.~O. Silva, D.~Gerosa, P.~Pani, E.~Berti, L.~Gualtieri, and
  U.~Sperhake, ``{Tensor-multi-scalar theories: relativistic stars and 3 + 1
  decomposition},''
  \href{http://dx.doi.org/10.1088/0264-9381/32/20/204001}{{\em Class. Quant.
  Grav.} {\bfseries 32} no.~20, (2015) 204001},
  \href{http://arxiv.org/abs/1505.07462}{{\ttfamily arXiv:1505.07462 [gr-qc]}}.

\bibitem{Yazadjiev:2019oul}
S.~S. Yazadjiev and D.~D. Doneva, ``{Dark compact objects in massive
  tensor-multi-scalar theories of gravity},''
  \href{http://dx.doi.org/10.1103/PhysRevD.99.084011}{{\em Phys. Rev. D}
  {\bfseries 99} no.~8, (2019) 084011},
  \href{http://arxiv.org/abs/1901.06379}{{\ttfamily arXiv:1901.06379 [gr-qc]}}.

\bibitem{Regge:1957td}
T.~Regge and J.~A. Wheeler, ``{Stability of a Schwarzschild singularity},''
\href{http://dx.doi.org/10.1103/PhysRev.108.1063}{{\em Phys. Rev.} {\bfseries
  108} (1957) 1063--1069}.

\bibitem{Zerilli:1971wd}
F.~J. Zerilli, ``{Gravitational field of a particle falling in a schwarzschild
  geometry analyzed in tensor harmonics},''
\href{http://dx.doi.org/10.1103/PhysRevD.2.2141}{{\em Phys. Rev.} {\bfseries
  D2} (1970) 2141--2160}.

\bibitem{Zerilli:1970se}
F.~J. Zerilli, ``{Effective potential for even parity Regge-Wheeler
  gravitational perturbation equations},''
\href{http://dx.doi.org/10.1103/PhysRevLett.24.737}{{\em Phys. Rev. Lett.}
  {\bfseries 24} (1970) 737--738}.

\bibitem{Sago:2002fe}
N.~Sago, H.~Nakano, and M.~Sasaki, ``{Gauge problem in the gravitational
  selfforce. 1. Harmonic gauge approach in the Schwarzschild background},''
  \href{http://dx.doi.org/10.1103/PhysRevD.67.104017}{{\em Phys. Rev.}
  {\bfseries D67} (2003) 104017},
\href{http://arxiv.org/abs/gr-qc/0208060}{{\ttfamily arXiv:gr-qc/0208060
  [gr-qc]}}.

\bibitem{Martel:2003jj}
K.~Martel, ``{Gravitational wave forms from a point particle orbiting a
  Schwarzschild black hole},''
  \href{http://dx.doi.org/10.1103/PhysRevD.69.044025}{{\em Phys. Rev.}
  {\bfseries D69} (2004) 044025},
\href{http://arxiv.org/abs/gr-qc/0311017}{{\ttfamily arXiv:gr-qc/0311017
  [gr-qc]}}.

\bibitem{Blazquez-Salcedo:2016enn}
J.~L. Blázquez-Salcedo, C.~F.~B. Macedo, V.~Cardoso, V.~Ferrari, L.~Gualtieri,
  F.~S. Khoo, J.~Kunz, and P.~Pani, ``{Perturbed black holes in
  Einstein-dilaton-Gauss-Bonnet gravity: Stability, ringdown, and
  gravitational-wave emission},''
  \href{http://dx.doi.org/10.1103/PhysRevD.94.104024}{{\em Phys. Rev.}
  {\bfseries D94} no.~10, (2016) 104024},
\href{http://arxiv.org/abs/1609.01286}{{\ttfamily arXiv:1609.01286 [gr-qc]}}.

\bibitem{degeneracies}
A.~Maselli and et~al.

\bibitem{Pani:2009wy}
P.~Pani and V.~Cardoso, ``{Are black holes in alternative theories serious
  astrophysical candidates? The Case for Einstein-Dilaton-Gauss-Bonnet black
  holes},'' \href{http://dx.doi.org/10.1103/PhysRevD.79.084031}{{\em Phys.
  Rev.} {\bfseries D79} (2009) 084031},
\href{http://arxiv.org/abs/0902.1569}{{\ttfamily arXiv:0902.1569 [gr-qc]}}.

\bibitem{Julie:2019sab}
F.-L. Juli\'e and E.~Berti, ``{Post-Newtonian dynamics and black hole
  thermodynamics in Einstein-scalar-Gauss-Bonnet gravity},''
  \href{http://dx.doi.org/10.1103/PhysRevD.100.104061}{{\em Phys. Rev.}
  {\bfseries D100} no.~10, (2019) 104061},
\href{http://arxiv.org/abs/1909.05258}{{\ttfamily arXiv:1909.05258 [gr-qc]}}.

\bibitem{Glampedakis:2005cf}
K.~Glampedakis and S.~Babak, ``{Mapping spacetimes with LISA: Inspiral of a
  test-body in a `quasi-Kerr' field},''
  \href{http://dx.doi.org/10.1088/0264-9381/23/12/013}{{\em Class. Quant.
  Grav.} {\bfseries 23} (2006) 4167--4188},
  \href{http://arxiv.org/abs/gr-qc/0510057}{{\ttfamily arXiv:gr-qc/0510057}}.

\bibitem{Barack:2006pq}
L.~Barack and C.~Cutler, ``{Using LISA EMRI sources to test off-Kerr deviations
  in the geometry of massive black holes},''
  \href{http://dx.doi.org/10.1103/PhysRevD.75.042003}{{\em Phys. Rev. D}
  {\bfseries 75} (2007) 042003},
  \href{http://arxiv.org/abs/gr-qc/0612029}{{\ttfamily arXiv:gr-qc/0612029}}.

\bibitem{Babak:2017tow}
S.~Babak, J.~Gair, A.~Sesana, E.~Barausse, C.~F. Sopuerta, C.~P. Berry,
  E.~Berti, P.~Amaro-Seoane, A.~Petiteau, and A.~Klein, ``{Science with the
  space-based interferometer LISA. V: Extreme mass-ratio inspirals},''
  \href{http://dx.doi.org/10.1103/PhysRevD.95.103012}{{\em Phys. Rev. D}
  {\bfseries 95} no.~10, (2017) 103012},
  \href{http://arxiv.org/abs/1703.09722}{{\ttfamily arXiv:1703.09722 [gr-qc]}}.

\bibitem{Cardoso:2018zhm}
V.~Cardoso, G.~Castro, and A.~Maselli, ``{Gravitational waves in massive
  gravity theories: waveforms, fluxes and constraints from extreme-mass-ratio
  mergers},'' \href{http://dx.doi.org/10.1103/PhysRevLett.121.251103}{{\em
  Phys. Rev. Lett.} {\bfseries 121} no.~25, (2018) 251103},
  \href{http://arxiv.org/abs/1809.00673}{{\ttfamily arXiv:1809.00673 [gr-qc]}}.

\bibitem{Piovano:2020ooe}
G.~A. Piovano, A.~Maselli, and P.~Pani, ``{Model independent tests of the Kerr
  bound with extreme mass ratio inspirals},''
  \href{http://arxiv.org/abs/2003.08448}{{\ttfamily arXiv:2003.08448 [gr-qc]}}.

\bibitem{Berti:2015itd}
E.~Berti {\em et~al.}, ``{Testing General Relativity with Present and Future
  Astrophysical Observations},''
  \href{http://dx.doi.org/10.1088/0264-9381/32/24/243001}{{\em Class. Quant.
  Grav.} {\bfseries 32} (2015) 243001},
\href{http://arxiv.org/abs/1501.07274}{{\ttfamily arXiv:1501.07274 [gr-qc]}}.

\bibitem{Pound:2015tma}
A.~Pound, ``{Motion of small objects in curved spacetimes: An introduction to
  gravitational self-force},''
  \href{http://dx.doi.org/10.1007/978-3-319-18335-0\_13}{{\em Fund. Theor.
  Phys.} {\bfseries 179} (2015) 399--486},
  \href{http://arxiv.org/abs/1506.06245}{{\ttfamily arXiv:1506.06245 [gr-qc]}}.

\bibitem{Barack:2018yvs}
L.~Barack and A.~Pound, ``{Self-force and radiation reaction in general
  relativity},'' \href{http://dx.doi.org/10.1088/1361-6633/aae552}{{\em Rept.
  Prog. Phys.} {\bfseries 82} no.~1, (2019) 016904},
  \href{http://arxiv.org/abs/1805.10385}{{\ttfamily arXiv:1805.10385 [gr-qc]}}.

\bibitem{Pound:2019lzj}
A.~Pound, B.~Wardell, N.~Warburton, and J.~Miller, ``{Second-Order Self-Force
  Calculation of Gravitational Binding Energy in Compact Binaries},''
  \href{http://dx.doi.org/10.1103/PhysRevLett.124.021101}{{\em Phys. Rev.
  Lett.} {\bfseries 124} no.~2, (2020) 021101},
  \href{http://arxiv.org/abs/1908.07419}{{\ttfamily arXiv:1908.07419 [gr-qc]}}.

\end{thebibliography}\endgroup

\end{document}